\documentclass[%
 reprint,
nofootinbib,
nobalancelastpage,
pra,
aps,
]{revtex4-2}
\usepackage{braket}
\usepackage{float} 
\usepackage{mhchem}
\usepackage{siunitx}
\usepackage{graphicx}
\usepackage{dcolumn}
\usepackage{bm}
\usepackage[dvipsnames]{xcolor}
\usepackage[colorlinks=true,citecolor=NavyBlue,urlcolor=NavyBlue]{hyperref}
\usepackage{booktabs}
\usepackage{orcidlink}
\usepackage[super]{nth}
\usepackage{dsfont}
\newcommand\blfootnote[1]{%
\begin{NoHyper}
  \begingroup
  \renewcommand\thefootnote{}\footnote{#1}%
  \addtocounter{footnote}{-1}%
  \endgroup
\end{NoHyper}  
}
\usepackage{algorithmicx}
\usepackage{algorithm}
\usepackage[noend]{algpseudocode}
\usepackage{bbm}
\usepackage{amssymb}
\usepackage{amsmath}
\DeclareMathOperator{\sgn}{sgn}
\usepackage{mathtools}
\DeclarePairedDelimiter{\abs}{\lvert}{\rvert}%
\DeclarePairedDelimiter\norm{\lVert}{\rVert}%
\makeatletter
\let\oldabs\abs
\def\abs{\@ifstar{\oldabs}{\oldabs*}}
\let\oldnorm\norm
\def\norm{\@ifstar{\oldnorm}{\oldnorm*}}
\makeatother

\DeclareMathOperator{\arccosh}{arccosh}

\usepackage[normalem]{ulem}

\begin{document}

\title{Benchmarking a randomized quantum algorithm for estimating expectation values of molecular observables}%

\date{\today}

\author{Anna Schroeder\orcidlink{0009-0006-4971-3265}$^{1}$}
\email[Corresponding author: ]{anna.schroeder@merckgroup.com} 
\affiliation{QUTAC Material Science Working Group}

\author{Gian-Luca R. Anselmetti\orcidlink{0000-0002-8073-3567}$^{2}$}
\affiliation{QUTAC Material Science Working Group}

\author{Matthias Degroote\orcidlink{0000-0001-5645-4667}$^{2}$}
\affiliation{QUTAC Material Science Working Group}

\author{Florian Dommert\orcidlink{0000-0001-7455-9702}$^{3}$}
\affiliation{QUTAC Material Science Working Group}

\author{Michael Kaicher\orcidlink{0000-0001-7986-5127}$^{4}$}
\affiliation{QUTAC Material Science Working Group}

\author{Chandan Kumar\orcidlink{0000-0001-6510-4204}$^{5}$}
\affiliation{QUTAC Material Science Working Group}

\author{Michael Streif\orcidlink{0000-0002-7509-4748}$^{2}$}
\email[Corresponding author: ]{michael.streif@boehringer-ingelheim.com} 
\affiliation{QUTAC Material Science Working Group}

\author{Davide Vodola\orcidlink{0000-0003-0880-3548}$^{4}$}
\affiliation{QUTAC Material Science Working Group}
\begin{abstract}
Randomized quantum algorithms for the estimation of expectation values of molecular observables offer a promising route to early fault-tolerant quantum advantage by trading circuit depth for an increased number of samples. However, this begs the question whether the resulting measurement overhead remains feasible for industrially relevant molecular systems. In this work, we benchmark the randomized sampling algorithm of Wang et al.~\cite{wang2024qubit} through numerical simulations on a hydrogen molecule and resource estimates for various active spaces of the cytochrome P450, an industry-relevant system in the pharmaceutical industry. Our numerical simulations reveal that the analytical error bounds significantly overestimate the actual sampling error, whereas the analytical bound on the required imaginary evolution time is comparatively tight, exceeding a direct DMRG estimate by roughly $44\%$ for the active space where the comparison is feasible. The measurement overhead nevertheless remains prohibitively high: even for the smallest active spaces, the expectation-value estimation alone requires $\sim 10^{10}$ samples, scaling to beyond $10^{15}$ for the largest active spaces considered. These results suggest that substantial improvements in measurement strategies or algorithmic scaling are necessary before randomized early-fault-tolerant algorithms become competitive for industrial quantum chemistry applications.
\end{abstract}

\maketitle

\section{Introduction }
\blfootnote{
$^1$ Merck KGaA, Frankfurter Stra{\ss}e 250, 64293 Darmstadt, Germany\\
$^2$ Quantum Lab, Boehringer Ingelheim, Ingelheim am Rhein, Germany\\
$^3$ TRUMPF SE + Co. KG, Johann-Maus-Str. 2, 71254 Ditzingen, Germany\\
$^4$ BASF Digital Solutions GmbH, Next Generation Computing, Pfalzgrafenstr. 1, 67056, Ludwigshafen, Germany\\
$^5$ BMW Group, Munich, Germany\\
}
Quantum computing promises to accelerate challenging computations in material science and quantum chemistry beyond the reach of classical supercomputers. However, current quantum devices remain far from the scale required to tackle classically intractable problems using fully fault-tolerant algorithms such as quantum phase estimation (QPE). To bridge this gap, recent research has focused on early-fault-tolerant quantum algorithms that reduce circuit-depth requirements at the cost of increased sampling overhead~\cite{campbell2019random,wan2022randomized,lin2022heisenberg,dong2022groundstate}. A central open question for these approaches is whether the resulting measurement overhead remains feasible for systems of practical interest. 

Randomized algorithms represent a particularly promising class of early-fault-tolerant methods due to their favorable asymptotic scalings and minimal qubit overhead. Wang et al.~\cite{wang2024qubit} introduced a generalized framework that enables randomized evaluation of arbitrary matrix functions by combining approximate linear combination of unitaries (LCU) decompositions~\cite{childs2021theory} with the Pauli-rotation decomposition scheme of Wan et al.~\cite{wan2022randomized}. When applied to the estimation of expectation values of molecular observables, this framework allows the user to obtain gate complexities comparable to state-of-the-art methods, but requiring only a single auxiliary qubit \cite{wang2024qubit}. However, asymptotic scalings alone are insufficient to judge the practical competitiveness of an algorithm. Constant-factor analyses with system-specific parameters are essential to determine whether a specific quantum algorithm can be executed on early-fault tolerant hardware for realistic problem sizes~\cite{hoefler2023disentangling, gunther2026phase,hariprakash2025randomized, oumarou2026optimizing, nie2026pauli, kiss2025early, wang2025randomized}.

In this work, we benchmark Wang’s randomized algorithm~\cite{wang2024qubit}, which we refer to as the randomized ground-state property sampling algorithm (RGPS), on molecular systems relevant to industrial quantum chemistry. Specifically, we (i) perform numerical simulations of the complete algorithm for a hydrogen molecule to assess its convergence behavior and the tightness of the analytical error bounds, and (ii) provide resource estimates for different active spaces of cytochrome P450, an industrially relevant system in the pharmaceutical industry, using input parameters obtained from density-matrix renormalization group (DMRG) calculations. Our analysis quantifies the measurement overhead as a function of system size and identifies the dominant cost drivers, thereby providing a concrete assessment of the algorithm's practicality for industrial applications.

\section{Observables in quantum chemistry}

While computing ground state energies is currently the main research direction of fault-tolerant quantum algorithms for quantum chemistry, many chemically relevant properties, such as forces, dipole moments, and response functions, require the evaluation of expectation values of observables that do not commute with the electronic structure Hamiltonian $H$ with ground state $\ket{\Phi_0}$.
For example, simulating molecular dynamics, i.e., the time evolution of nuclei within a molecule, requires the evaluation of forces acting on the nuclei. One approach to computing the forces is by taking the derivative of the electronic energy $E$ with respect to a nuclear coordinate: 
\begin{align*}
	\braket{F_a^{(i)}} = -\frac{\mathrm{d}E}{\mathrm{d}R_a^{(i)}}\,,
\end{align*} 
where $R_a^{(i)}$ denotes the i-th cartesian component of the position of nucleus a, with $i\in\{x,y,z\}$. Alternatively, the force can be obtained by evaluating the expectation value of the force operator with respect to the ground state of the electronic Hamiltonian: 
\begin{align*}
	\braket{F_a^{(i)}} = \braket{\Phi_0|F_a^{(i)}|\Phi_0}\,,
\end{align*} 
Here, $\ket{\Phi_0}$ is the ground state of the electronic structure Hamiltonian and $F_a^{(i)}$ is the force operator given by 
\begin{align*}
F_{a}^{(i)} = \sum_{pq} f_{pq}^{(1)} a_{p}^{\dagger} a_{q} + \sum_{pqrs} f_{pqrs}^{(2)} a_{p}^{\dagger} a_{q}^{\dagger} a_{r} a_{s}\,,
\end{align*}
where $f^{(1)}_{pq}$ and $f^{(2)}_{pqrs}$ are one- and two-body integrals, as defined in Ref.~\cite{obrien2022efficient}.
Although $F_a^{(i)}$ shares the second-quantized structure of the electronic structure Hamiltonian, its coefficients differ in origin. Since neither the kinetic energy nor the electron--electron interaction depends on the nuclear coordinates, differentiating at fixed basis yields a purely one-body operator. The two-body coefficients $f^{(2)}_{pqrs}$ arise from the nuclear-coordinate dependence of atom-centered basis functions: they contain the derivatives of the two-electron integrals with respect to $R_a^{(i)}$. For a basis independent of the nuclear positions, such as plane waves or a real-space grid, $f^{(2)}$ vanishes.

Since the force operators do not commute with the Hamiltonian, i.e. $[H,F_a^{(i)}]\neq 0$, their expectation values differ from eigenvalues and are not directly computable via quantum phase estimation (QPE). Instead, special fault-tolerant quantum algorithms have been designed for this task in Refs.~\cite{knill2007optimal,steudtner2023fault}, albeit incurring higher costs than ground state energy estimation alone. 
Forces are thus the canonical example of a chemically relevant observable that requires a dedicated expectation-value estimation protocol rather than QPE, and we use them throughout as the motivating target for the randomized measurement protocol introduced by Wang et al.~\cite{wang2024qubit} for early-fault-tolerant quantum computers. We stress at the outset that the resource estimates of Sec.~\ref{sec:resource_estimates} are carried out for the expectation value of the active-space Hamiltonian rather than for an individual force component. As discussed there, the dominant cost drivers---initial-state overlap, spectral gap, and imaginary time---are observable independent, so the two targets differ only through the operator norm entering the sampling bound.

\section{Randomized algorithm for expectation value estimation}

\label{sec:theory}
In this section, we provide an overview of RGPS introduced by Wang et al. in Ref.~\cite{wang2024qubit}. RGPS is a concrete instance of a more general framework developed in the same work, which enables randomized evaluation of expressions such as
$\bra{\phi} s(A) \ket{\psi}$ or $\mathrm{Tr}(s(A) \rho\, s(A)^\dagger O)$ where $\ket{\phi}, \ket{\psi}$ are pure quantum states and $\rho$ a general state prepared on a quantum computer, and $s(A)$ is a Fourier series approximation to a matrix function $f(A)$
\begin{equation}
    f(A) \approx s(A) = \sum_{k \in F} \alpha_k \exp(i t_k A),
\end{equation} with $\alpha_k$ the coefficient associated to the $k$-th Fourier term and  $A$ given in the Pauli-access model 
\begin{equation}
    A = \sum_\ell a_\ell P_\ell,
\end{equation} decomposed into so-called Pauli strings $P_\ell$ and suitable coefficient $a_\ell$. The key idea is to sample from an ensemble of quantum circuits composed of Clifford gates and a user-defined number of non-Clifford gates per circuit, using only a single auxiliary qubit. We first review the scientific developments that led to the Wang scheme and then discuss how it can be applied to ground-state property sampling. As will become clear, RGPS is a combination of a Fourier approximation and two sampling steps, both of which rely on the validity of their formal assumptions. Here, we aim to explain these ingredients for RGPS in the context of the electronic-structure Hamiltonian.

Ref.~\cite{wang2024qubit} presents a generalization of the work of Wan et al.~\cite{wan2022randomized}, who proposed a ``double-randomized'' quantum phase estimation protocol. This protocol combines the qubit-efficient randomized quantum phase estimation method of Lin and Tong~\cite{lin2022heisenberg} with a second, distinct randomization technique called randomized compilation. In the approach of Lin and Tong, one extracts the eigenvalue associated with a state of interest by estimating an approximate cumulative distribution function (CDF) of the Hamiltonian spectrum. This is achieved by expressing the approximate CDF as a Fourier series. A favorable runtime is then obtained by not estimating every term in the series individually, but instead applying importance sampling to the series as a whole. The sample complexity was shown to scale as $\mathcal{O}(\alpha^2/\sigma^2)$, where $\alpha = \sum_{k \in F} |\alpha_k|$ is the $\ell_1$ norm of the Fourier series, and $\sigma^2$ is the target variance of the estimator. This constitutes the first randomization technique relevant when the problem is formulated in terms of a Fourier series.
The second randomization concerns the compilation of each Fourier term. Wan et al. proposed a randomized compilation scheme (Lemma 2 of Ref.~\cite{wan2022randomized}) that allows to decompose any Hamiltonian evolution operator as 
\begin{equation}\label{eq:itdecomp}
    e^{i t \hat{H}} = \sum_k b_k^{(r)} U_k^{(r)},
\end{equation}
where $\hat{H}$ is a Hermitian operator written as a normalized linear combination of Pauli operators, and the $b_k^{(r)} \in \mathbb{R}$ are coefficients whose normalized absolute values $|b_k^{(r)}|/\sum_{k'}|b_{k'}^{(r)}|$ serve as the sampling probability of the unitary $U_k^{(r)}$, with the sign $\sgn(b_k^{(r)})$ carried into the estimator. Each $U_k^{(r)}$ consists of a sequence of Clifford operations and $r$ non-Clifford operations in total. In terms of implementation cost, this user-defined number of non-Clifford operations, $r$, therefore serves as a proxy for the circuit depth of each sampled circuit.
The explicit form of $b_k$ and $U_k$ follows from the derivation in Appendix C of Ref.~\cite{wan2022randomized}
\begin{subequations}\label{eq:randomizedcomp}
\begin{align}
    e^{i\hat{H}x}
    &= \sum_{n \text{ even}} \frac{(ix)^n}{n!} \sqrt{1 + \left(\frac{x}{n+1}\right)^2} \nonumber\\
    &\quad\times \sum_{\ell_1, \dots, \ell_n, \ell'} p_{\ell_1}\cdots p_{\ell_n} p_{\ell'}
       \left(P_{\ell_1}\cdots P_{\ell_n} V_{\ell'}^{(n)} \right) \\
    &=: \sum_m c_m W_m ,\label{eq:lcu1}
\end{align}
\end{subequations}
where $x \coloneqq t/r$ is the evolution time per non-Clifford segment and
\begin{align}\label{eq:angles}
    V_{\ell'}^{(n)} &\coloneqq \frac{1}{\sqrt{1 + \left(\frac{x}{n+1}\right)^2}} \left(\mathbbm{1} + i\frac{x}{n+1} P_{\ell'}\right)= \exp(i\theta_n P_{\ell'}), \\ 
    &\quad\text{with }\theta_n \coloneqq \arccos\left(\left[1 + \left(\frac{x}{n+1}\right)^2 \right]^{-1/2}\right).
\end{align}
Here, $p_\ell = a_\ell / \lambda$ are the normalized Pauli weights of the Pauli string $P_\ell$ appearing in the Hamiltonian, with $\lambda = \sum_\ell |a_\ell|$. Note that the $p_\ell$, and consequently the coefficients $c_m$ and $b_k$, are real but not necessarily positive; sampling is performed according to $|p_\ell|$ and the signs are absorbed into the estimator. To implement the full time-evolution operator, one must multiply $r$ such terms. Thus, $U_k = W_{m_1} \cdots W_{m_r}$. Note, that sampled operators rarely contain Pauli operators $P_{\ell_i}$ since the coefficients decay factorially with $n$. The sampling complexity of the resulting importance sampling procedure depends on the total weight of the LCU that is determined via the evolution time $t$ and the chosen depth $r$. Proposition 13 of Ref.~\cite{wan2022randomized} shows that this quantity can be bounded as
\begin{align}
    \sum_k |b_k| = (\sum_m |c_m|)^r \leq e^{t^2/r}.
\end{align}
This provides an important intuition that will also be relevant for the use-case-specific benchmarking discussed in Sec.~\ref{sec:resource_estimates}; since the randomized compilation relies on the Taylor expansion of $\exp(i t \hat{H}/r)$, one can only yield an precise approximation to the Hamiltonian evolution operator with tractable resources when $t^2/r$ remains at most of constant order. In particular, one cannot choose $r$ arbitrarily small and still obtain an efficient shallow-depth algorithm in terms of sampling costs.

While the results of Ref.~\cite{wan2022randomized} were derived for a particular Fourier series representation of a matrix function, the CDF of the spectrum, the underlying derivations can extend to general Fourier series. This generalization was developed by Wang et al.~\cite{wang2024qubit}, who provided rigorous resource and performance guarantees for randomized evaluation of expressions such as
$\bra{\phi} f(A) \ket{\psi}$ or $\mathrm{Tr}(f(A)^\dagger \rho f(A) O)$ given a sufficiently close Fourier approximation $s(A)$ to $f(A)$. They further studied three specific use cases for which previous work had already established well-converging Fourier approximations for relevant matrix functions: matrix inversion, Gibbs-state sampling, and ground-state property sampling.

The latter application was previously studied by Keen et al.~\cite{keen2021quantum} in the context of early fault-tolerant Green’s-function estimation. The matrix function of interest is the imaginary-time evolution (ITE) operator, $\exp(-\frac{1}{2} H^2\tau_s^2 )$, which projects an initial state $\ket{\psi}$ onto the ground state of the Hamiltonian $H$, provided that $\tau_s$ is sufficiently large and $\braket{\Phi_0|\psi}\neq 0$, where $\ket{\Phi_0}$ denotes the ground state. Importantly, Keen et al. consider the squared Hamiltonian $H^2$ in order to exploit the Hubbard--Stratonovich transformation~\cite{keen2021quantum},
\begin{align}\label{eq:ITEexact}
f(H) = e^{-\frac{1}{2} H^2\tau_s^2} = \frac{1}{\sqrt{2\pi}} \int dz \, e^{-\frac{1}{2} z^2} e^{-i z H \tau_s}.
\end{align}
However, this choice implies that the ITE operator converges to the eigenstate of $H$ with the smallest absolute eigenvalue, which may differ from the ground state, particularly for nonpositive Hamiltonians such as the electronic structure Hamiltonian. A simple remedy is to shift the spectrum by a sufficiently large constant, for example using a lower bound $E_{0} \geq E_0^\mathrm{lb}$ to the ground-state energy.
Lemma 1 in~\cite{keen2021quantum} specifies a sufficient choice of $\tau_s$. Consider a Hamiltonian $H$ with spectrum $\sigma(H)\subseteq(0,1)$ and spectral gap $\Delta_s\geq\Delta_\ell>0$. Throughout this work, $\Delta_s$ denotes the exact gap of the rescaled Hamiltonian, so that $\Delta_s=\Delta/\mathcal{N}$ in terms of the unscaled gap $\Delta$ in Ha and the normalization $\mathcal{N}$ introduced below, and $\Delta_\ell$ denotes a known lower bound on $\Delta_s$. Let $\ket{\Phi_0}$ be the ground state with eigenvalue $E_0\geq 0$, and let $\ket{\psi_0}$ be an initial state with known overlap lower bound $|\braket{\Phi_0|\psi_0}|\geq\gamma>0$. Then the infidelity of the  normalized state $\ket{\psi} = f(H)\ket{\psi_0}/ \norm{ f(H)\ket{\psi_0}}$ with respect to the actual ground state satisfies 
\begin{align}\label{eq:keentimefidel} \textstyle
       1 - \abs{\braket{\Phi_0|\psi}} 
  \leq \frac{1}{2}\norm{\ket{\psi} - \ket{\Phi_0}}^2
  \leq \frac{\eta^2}{2} \equiv \varepsilon_\mathrm{state},
\end{align} if 
\begin{align} \textstyle
  \tau_s \geq \frac{1}{\Delta_\ell}\sqrt{2\log\frac{1}{\gamma\eta}},
\end{align} 
where $\eta \geq \norm{\ket{\psi} - \ket{\Phi_0}}$ bounds the Euclidean distance between the states. This lemma assumes that the spectrum is normalized to the unit interval, which can be achieved by rescaling the Hamiltonian by an upper estimate $\mathcal{N}$ of the spectral width; throughout this work we take $\mathcal{N}=2\lambda$, with $\lambda$ the Pauli one-norm defined below.

In summary, the electronic-structure Hamiltonian $H_0$, with ground-state energy $E_0$ and Pauli one-norm $\lambda=\sum_\ell|a_\ell|$ (excluding the identity contribution), must be preprocessed in two independent steps. We first shift it,
\begin{equation}\label{eq:shift}
    H' \coloneqq H_0 + \Lambda\,\mathbbm{1}\,,
    \qquad \Lambda \coloneqq -E_0^\mathrm{lb}\,,
\end{equation}
where $E_0^\mathrm{lb}\leq E_0$ is a classical lower bound on the ground-state energy, so that $H'\succeq0$. The residual ground-state energy after shifting,
\begin{equation}\label{eq:residual}
    \lambda_0 \coloneqq E_0 + \Lambda = E_0 - E_0^\mathrm{lb} \;\geq\; 0\,,
\end{equation}
is therefore precisely the tightness of that bound. We then normalise,
\begin{equation}\label{eq:preprocessing}
    \tilde H \coloneqq \frac{H'}{\mathcal{N}} = \frac{H_0+\Lambda\,\mathbbm{1}}{\mathcal{N}}\,,
    \qquad \mathcal{N} \geq E_{\max}+\Lambda\,,
\end{equation}
which gives $\sigma(\tilde H)\subseteq[0,1]$. Note that $\Lambda$ and $\mathcal{N}$ are independent choices: $\Lambda$ is fixed by the quality of the available lower bound on $E_0$, whereas $\mathcal{N}$ need only exceed the width of the shifted spectrum. Since $\sigma(H_0)\subseteq[c-\lambda,c+\lambda]$, with $c$ the identity coefficient of the Pauli decomposition, one has $E_{\max}+\Lambda\leq 2\lambda+\lambda_0$; we therefore fix $\mathcal{N}=2\lambda$ throughout, which is sufficient whenever $\lambda_0\ll\lambda$, as is the case for all systems considered here.

To keep the notation light, we write $H$ for the preprocessed Hamiltonian $\tilde H$ throughout this section, and $\tau_s$ for the imaginary time expressed in the corresponding rescaled units. The physical imaginary time in Ha$^{-1}$, which is the quantity reported in Sec.~\ref{sec:resource_estimates} and Tab.~\ref{tab:bound_parameters}, is $\tau = \tau_s/\mathcal{N}$.

Two additional remarks are in order. First, guaranteeing convergence to the true ground state requires a lower bound on the spectral gap, which may be obtained, for example, from an upper bound on the ground-state energy and a lower bound on the first excited-state energy, which is also a common prerequisite for many ground-state preparation algorithms \cite{ge2019faster}. While the former is well studied within variational reduced-density-matrix theory \cite{coleman1963structure, mazziotti2002variational, mazziotti2004realization, deprince2024variational}, methods for rigorously lower bounding excited-state energies are, to our knowledge, not well established. This also appears to hold within variational reduced-density-matrix theory, which is the main framework concerned with lower bounds on ground-state energies of electronic structure Hamiltonians; analogous methods for excited states are not directly available and remain an active area of research \cite{deprince2024variational}. Second, the method requires a lower bound on the overlap with the true ground state.
In our benchmarks, we address the lack of rigorous access to these quantities by restricting ourselves to system sizes that remain tractable with higher-level methods such as DMRG. This provides estimates of excited state energies and an accurate approximation to the ground state wavefunction, which we use as a surrogate for the true ground state when evaluating the overlap with, e.g., the Hartree--Fock state. Since these quantities are not rigorous lower bounds, the resulting procedure remains heuristic in practice.

Keen et al.~\cite{keen2021quantum} developed a Fourier approximation to the ITE operator in Eq.~\eqref{eq:ITEexact}, with bounded error relative to the exact operator. This is the content of Theorem 1 in Ref.~\cite{keen2021quantum}, which assumes the same conditions as Lemma 1 (cf. Eq.~\ref{eq:keentimefidel}). In particular,
\begin{equation}\label{eq:ITEfourier}
    \norm{s(H) - f(H)} \leq \gamma \eta \equiv \varepsilon_\mathrm{op},
\end{equation} where the error is measured in spectral norm. This bound is obtained by discretizing and truncating the integral in Eq.~\eqref{eq:ITEexact},
\begin{align} \label{eq:discretizedHS}
    s(H) &= \frac{1}{\sqrt{2\pi}} \sum_{k=-N_z}^{N_z}\Delta_z
     e^{ -\frac{1}{2}z_{k}^2} e^{-i z_{k} \tau_s H } \\
  &\equiv \sum_{k=-N_z}^{N_z} \alpha_k U_k,
  \quad (z_k = k\Delta_z).
\end{align}
and by separately bounding the discretization and truncation errors, assuming that both contribute equally. This analysis is given in Appendix A.2. of Ref.~\cite{keen2021quantum}. Note that this approximation step introduces a bias that cannot be compensated for by concurrent steps. Under the same assumptions, Keen et al. further show that, analogously to Eq.~\eqref{eq:keentimefidel}, the infidelity of the projected state produced by the approximate ITE operator, which we indicate as $\mathrm{tITE}$, satisfies
\begin{align}
   1 - \abs{\braket{\Phi_0|\psi}} 
  \leq \frac{1}{2}\norm{\ket{\psi} - \ket{\Phi_0}}^2
 \leq \frac{(5\eta)^2}{2} \equiv \varepsilon_\mathrm{state},
\end{align}
provided that 
\begin{align}
  \tau_s &\textstyle \geq \frac{1}{\Delta_\ell}\sqrt{2\log\frac{1}{\gamma\eta}}
  = \mathcal{O}(\frac{1}{\Delta_\ell}\sqrt{\log\frac{1}{\gamma\eta}}),\\
  \lambda_0^{(s)} &\leq \frac{1}{\tau_s}
  =  \mathcal{O}(\Delta_\ell / \sqrt{\log\frac{1}{\gamma\eta}}),
\displaybreak[1] \\
  z_c &\textstyle = N_z \Delta_z
  \geq \sqrt{2\log\frac{2}{\gamma\eta}}
  = \mathcal{O}(\sqrt{\log\frac{1}{\gamma\eta}}),
\displaybreak[1] \\  
   \Delta_z &\textstyle = \frac{2\pi}{z_c' + \tau_s}
  = \mathcal{O}(\Delta_\ell / \sqrt{\log\frac{1}{\gamma\eta}}), \\
  N_z &= \lceil\frac{z_c}{\Delta_z}\rceil
  = \mathcal{O}(\frac{1}{\Delta_\ell} \log \frac{1}{\gamma\eta}).
\end{align}
All quantities in this list refer to the rescaled Hamiltonian, with $\lambda_0^{(s)}=\lambda_0/\mathcal{N}$, and $z_c'$ is defined in Appendix~\ref{sec_param_estimate}.

Given a rigorous approximation to the ITE operator, we can construct the ground-state property estimator on which our RGPS is based:
\begin{align}
\label{eq:ITE}
    \braket{O}_{\mathrm{ITE}} = \frac{
    \braket{\psi_0 | e^{-\frac{1}{2} H^2\tau_s^2} O\, e^{-\frac{1}{2}H^2\tau_s^2} | \psi_0}}{\braket{\psi_0 | e^{-H^2\tau_s^2}| \psi_0}}\,.
\end{align}
Since the ITE operator is non-unitary, the normalization factor in the denominator must be estimated separately. Both numerator and denominator are of the form $\mathrm{Tr}(f(A)^\dagger \rho f(A) O)$ where $O$ in the numerator is the operator of interest for the studied system and $O=I$ for the denominator. 

What is still missing is a rigorous bound on the total error of the final estimator that combines the Fourier approximation with the two randomization steps, together with the corresponding sampling complexity and circuit depth per sample.
After applying the Fourier approximation and randomized compilation, the time-evolution operator takes the form
\begin{align}
    e^{-\frac{1}{2}H^2\tau_s^2} = \sum_{k,m} \alpha_k \beta_{km}^{r_k} U_{km}^{r_k},
\end{align} where $U_{k,m}^{r_k}$ has the form given in Eq.~\eqref{eq:lcu1}, $\alpha_k$ is the coefficient of the $k$-th term of the Fourier series, and $\beta_{km}^{r_k}$ the coefficient of the $m$-th term of the randomized compilation of the $k$-th Fourier term given a maximal ``depth'' $r_k$. For the ITE case, $k \in \{-N_z,\ldots,N_z\}$, while $m$ indexes the possible combinations of Pauli strings and rotations about their respective axes, as defined in Eq.~\ref{eq:randomizedcomp}; in practice, $m$ restricts to combinations with a rotation alone. Our goal is to approximate the expectation value of the imaginary-time-evolved initial state, which can be written as
\begin{align}
    \mathrm{Tr}&(s(A)\rho s(A)^\dagger O) \\ &=R(\vec{r})^2 \sum_{(k,m),(k',m')} p_{km}^{(r_{k})} p_{k'm'}^{(r_{k'})} \mathrm{Tr}(U_{km}^{(r_{k})} \rho U_{k'm'}^{(r_{k'})\dagger}),
\end{align} where
\begin{align}\label{eq:random_unitaries}
    R(\vec{r}) = \sum_{k,m} |\alpha_k \beta_{km}^{(r_k)}|,
\end{align} and
\begin{align}
    p_{km} = \frac{|\alpha_k \beta_{km}^{(r_k)}|}{R(\vec{r})}.
\end{align}
Accordingly, the left-hand side can be estimated by sampling, for each shot, two unitaries $U_{km}^{(r_{k})}$ and $U_{k'm'}^{(r_{k'})}$ and applying a measurement scheme such as the Hadamard test in the generalized form used for LCU-based simulation~\cite{childs2012hamiltonian}. Let the outcome of the (generalized) Hadamard test be $z_j \in [-\norm{O}R(\vec{r})^2,\norm{O}R(\vec{r})^2]$. Then Hoeffding’s inequality implies that the empirical estimator
\begin{align}
    \bar{z}^{(M)} = \frac{1}{M} \sum_j z_j,
\end{align}
\begin{align}
    |\bar{z}^{(M)} - \mathrm{Tr}(s(A) \rho s(A)^\dagger O)| \leq \varepsilon
\end{align} with success probability at least $1-\delta$, provided that the sampling count $M$ is at least
\begin{align}
    M \geq \log(\frac{2}{\delta}) \frac{2 \norm{O}^2 R(\vec{r})^4}{\varepsilon^2}.
\end{align} 
Since this statistical error assumes access to the Hadamard test result, the additional shot-noise scaling required to obtain that result appears only implicitly and should be included in a more realistic end-to-end resource estimate.
Including the error incurred by approximating the original matrix function $f(A)$ by its Fourier approximation, $|f(A)-s(A)|\leq \varepsilon$, yields Theorem 2 of Ref.~\cite{wang2024qubit}. Specifically, given a Fourier approximation satisfying
\begin{align}\label{eq:accuracyfourier}
    |f(A) - s(A)| \leq \frac{\varepsilon q^2}{6 \norm{O} \norm{f(A)}},
\end{align} one obtains an $\varepsilon$-accurate approximation to $\mathrm{Tr}(f(A) \rho f(A)^\dagger O)$ with the proposed randomized algorithm using the following resources
\begin{align}
   C_{\mathrm{sample}} &= \log(\frac{2}{\delta}) \frac{2 \norm{O}^2 R(\vec{r})^4}{\varepsilon^2 q^4} \\
    C_{\mathrm{depth}} &= 2 \lambda^2 t_{\mathrm{max}}^2 +d_\rho,
\end{align} 
where $R(\vec{r})$ and the maximal evolution time $t_\mathrm{max}$ in the Fourier series are assumed to satisfy the required Fourier-approximation accuracy in Eq.~\eqref{eq:accuracyfourier}, and $d_\rho$ denotes the circuit depth required to prepare the initial state. Note that this statement assumes access to the exact normalization factor $q$. Moreover, Wang showed in Proposition 2 (via Lemmas 3 and 4 in the appendix) that using an error tolerance three times smaller for estimating the normalization constant with the same randomized algorithm yields a sufficiently accurate estimate of $q$, so that Theorem 2 remains valid when $q$ is replaced by its approximation.

We leave with a final note on the preprocessing and resource adjustments required when the Hamiltonian is not positive. As already noted, to obtain an accurate Fourier approximation of the ITE operator of the electronic structure Hamiltonian, following the result of Keen et al., the Hamiltonian must be preprocessed to a) be positive, so that ITE converges to the desired ground state, and b) be normalized, in order to bound the error of the ITE operator. Wan et al.~\cite{wan2022randomized} and Wang et al. \cite{wang2024qubit} comment on nonpositive operators in the context of randomized compilation. Each Fourier term of the form $e^{i z_k H}$ can be randomly compiled when $H$ is preprocessed as a convex sum of Pauli strings, which effectively amounts to rescaling the imaginary time by $\lambda$: $e^{iz_k H} = e^{iz_k \lambda (H/\lambda)} $. As emphasized in \cite{wang2024qubit} and \cite{wan2022randomized}, shifting the Hamiltonian to make it nonnegative does not change the effective evolution time, and therefore does not change the resources required for randomized compilation since the constant shift can be factored out and absorbed into the Fourier coefficients, without affecting the normalization constant: $e^{iz_k (H+E_0)} = e^{iz_k \lambda (H/\lambda)}e^{iz_k E_0} $. Nevertheless, we emphasize that, in order to guarantee a Fourier approximation with certified fidelity, we must assume access to a nonnegative Hamiltonian with bounded spectrum, i.e., $H := (H_0 + E_0^\mathrm{lb})/(2\lambda)$ as introduced above. This raises the question of how such an input affects the resource requirements of randomized compilation $e^{iz_k \frac{H_0+E_0^\mathrm{lb}}{2\lambda }} = e^{iz_k \frac{\lambda}{\lambda + E_0^\mathrm{lb}} (\frac{H}{\lambda})}e^{iz_k \frac{E_0^\mathrm{lb}}{2\lambda }}$.

\section{Numerical simulations of a hydrogen molecule}
\label{sec:numerics}

In this section, we conduct numerical tests of the RGPS quantum algorithm on a simple molecular system, a hydrogen molecule ($\ce{H2}$) at a bond distance of $\SI{1}{\angstrom}$, represented in the minimal basis set STO-3G. The purpose of these simulations is to obtain a non-asymptotic estimate of the cost of the algorithm, complementing the analytical bounds derived in the previous section. We restrict ourselves to this minimal example because the number of circuit samples required to resolve the expectation values grows rapidly with system size, rendering a direct simulation of larger molecules or larger basis sets computationally infeasible. 

We start our numerical simulation by computing the imaginary time $\tau$ necessary to approximate the ground state to within an error in the overlap of $1 - \abs{\braket{\Phi_0|\psi}} \leq \varepsilon_\mathrm{state}=10^{-3}$. For this system, the necessary imaginary time duration can be calculated by explicitly constructing the ITE matrix $\mathrm{e}^{-H^2\tau^2/2}$ and subsequently applying it to the initial Hartree-Fock state $\ket{\mathrm{HF}}$.
After normalization of the resulting state, one can evaluate the overlap with the ground state to assess the accuracy of the approximation. This procedure requires access to the full spectrum of $H$ and is therefore only possible for small system sizes.

\begin{figure}[t]
\centering
\includegraphics[width=\columnwidth]{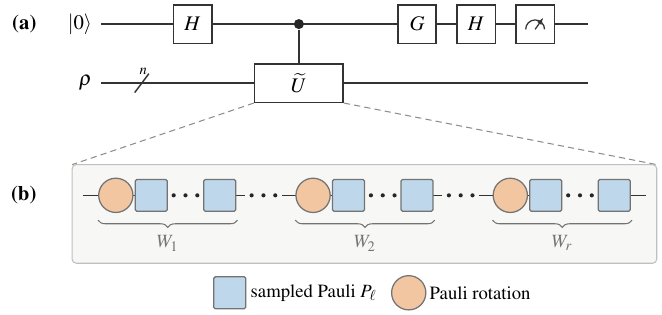}
\caption{Structure of the randomly compiled circuits developed in \cite{wan2022randomized}. (a)~ Hadamard test on the state $\rho$ and the randomly sampled unitary $\widetilde{U}$. Choosing $G=\mathbbm{1}$ and assigning the outcomes $\ket{0},\ket{1}$ the values $+1,-1$ yields an unbiased estimator of $\mathrm{Re}\,\mathrm{Tr}(\rho\widetilde{U})$; the choice $G=S^{\dagger}$ gives the corresponding imaginary part. (b)~Internal structure of $\widetilde{\mathcal{U}}$ for $H=\sum_{\ell}p_{\ell}P_{\ell}$. The circuit consists of $r$ segments $W_{i}$, each built from Pauli operators sampled according to $\{p_{\ell}\}_{\ell}$ interleaved with multi-qubit Pauli rotations. The number of Pauli operators within a given segment is itself random.}
\label{fig:depth_vs_samples}
\end{figure}

Following this, we utilize the computed imaginary time $\tau$ to determine the Fourier parameters as outlined in Appendix~C of Ref.~\cite{wan2022randomized} based on Eq.~\ref{eq:randomizedcomp}. Given these parameters, one can use the algorithm developed by Wang et al.~\cite{wan2022randomized} to compute the expectation value of observables of interest. We perform numerical simulations of the quantum algorithm over a range of values for the parameter $r$, which allows for a trade-off between the number of circuit samples and the depth of each circuit. Since $r$ is a run-time parameter, one can opt for different strategies, which we explore in Sec.~\ref{sec:resource_estimates}. Here, we keep $r$ fixed across all Fourier points. This choice is motivated by the fact that it represents the practically relevant scenario in which the available circuit depth is constrained by hardware limitations.
Specifically, a higher value of $r$ results in a deeper circuit, whereas a lower value of $r$ necessitates a greater number of circuit samples. This trade-off is illustrated in Fig.~\ref{fig:depth_vs_samples}. 
Concretely, we consider $r \in \{1, 10, 20, 100, 500, 2000\}$. In our simulation, for each value of $r$, we perform $10$ experiments, where for each  we draw $10^6$ circuit samples to estimate the expectation value of the targeted observable.

The numerical simulations evaluate the expectation value associated with each sampled randomized circuit exactly. Note, that they do not include the measurement-level variance arising from the generalized Hadamard test. While the analytical sample-complexity bounds already account for this measurement noise, the convergence observed in the numerical simulations should therefore be regarded as somewhat optimistic.

\begin{figure*}
  \includegraphics[width=0.999\textwidth]{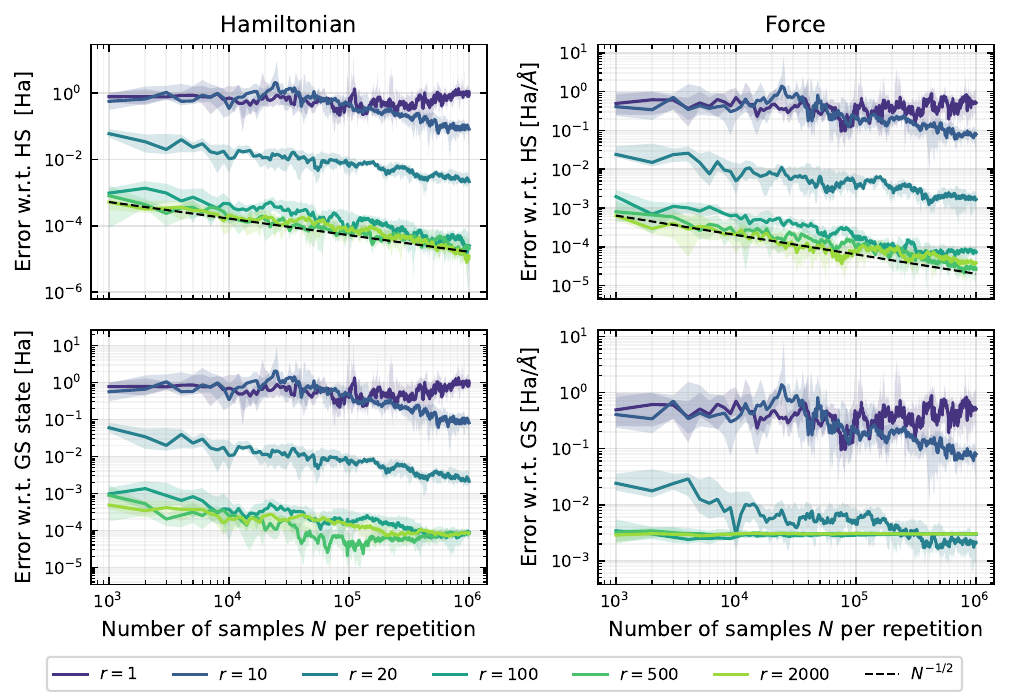}
  \caption{\label{fig:simulation_results} Simulation results with different observables. In the left column, the plots illustrate the outcomes when the Hamiltonian is used as the observable, while in the right column the plots depict the results for the force operator as the observable. In both instances, the top plot quantifies the error relative to the expectation value obtained from the truncated imaginary time evolved state via the truncated Hubbard-Stratonovich (HS) transformation, while the bottom plot computes the error with respect to the exact ground state. Each solid curve is the median over ten independent repetitions of $10^{6}$ samples, and the shaded band spans the interquartile range of those ten values at each $N$; the band is therefore the spread of individual runs, not an uncertainty on the median. Note that $N$ counts samples per repetition. The black dashed line in the top row is an $N^{-1/2}$ guide, anchored to the largest $r$ at the left edge.}
\end{figure*}

The computed expectation value then is compared to (i) the expectation value obtained w.r.t. the exact, but truncated imaginary-time evolved state $\braket{O}_\mathrm{tITE}$ and (ii) to the expectation value w.r.t. the exact ground state of the system $\braket{O}_\mathrm{GS}$. The results of our simulations are shown in Fig.~\ref{fig:simulation_results}. Specifically, Fig.~\ref{fig:simulation_results}~(left) displays the outcomes when the Hamiltonian serves as the observable of interest, $O=H$. Similarly, Fig.~\ref{fig:simulation_results}~(right) presents the results obtained when the observable is defined by the force operator $O=F$. 

In the top panel of Fig.~\ref{fig:simulation_results} , we observe that the error in estimating the expectation values of the Hamiltonian or the force operator with respect to the imaginary time-evolved state decreases both with the increase in the parameter $r$ and the number of samples. For very small values of $r$ (i.e., $r=1$), the error tends to fluctuate significantly, which aligns with the predictions of Eq.~(C4) of Ref.~\cite{wang2024qubit}. Conversely, at higher values of $r$, further increasing $r$ does not significantly enhance the rate of error convergence. In the bottom panel, the error in the sampled expectation values is compared against the expectation values derived from the exact ground state.
Given that the imaginary time was selected to generate a state with an infidelity of $10^{-3}$ relative to the ground state, the error in the expectation values reaches a plateau once it attains a certain threshold.

\begin{figure}[h]
\centering
\includegraphics[width=\columnwidth]{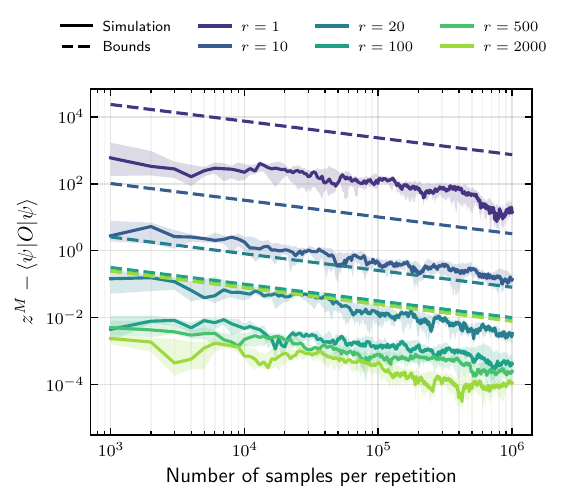}
\caption{\label{fig:simulations_vs_bounds}Comparison of the numerical simulation error with the analytical bound from Eq.~\eqref{eq:total_samples} for various values of the parameter $r$. Solid lines represent the simulation results, while dashed lines denote the bounds. Each color corresponds to a distinct value of $r$. Both bounds and simulation results pertain to the unnormalized expectation value of the observable. Each solid curve is the median over ten independent repetitions of $10^{6}$ samples and the shaded band their interquartile range, so $N$ counts samples per repetition. }
\end{figure}

While the numerical simulations provide direct estimates of the sampling cost for the small systems considered here, such simulations become infeasible for the larger systems studied in the following sections. To assess the sampling requirements in those cases, we derive in Appendix~\ref{app:optimization} an analytical upper bound on the estimation error. Before applying this bound to larger systems, we first benchmark it against the numerical results obtained here. The comparison is shown in Fig.~\ref{fig:simulations_vs_bounds}. We find that the bound reproduces the qualitative dependence on the filtering parameter $r$, although, as expected for a worst-case Hoeffding's bound, it overestimates the actual error observed in the simulations.

However, even these small-system simulations indicate that the number of circuit samples required can become substantial,  particularly for small values of $r$, even in the idealized scenario of having access to the exact expectation values, i.e. ignoring the additional overhead required to ensure each circuit sample converges within a reasonable statistical error by employing methods such as the Hadamard test.

\section{Resource estimates for larger systems}
\label{sec:resource_estimates}

From a practical standpoint, it is crucial to understand how the required number of samples and the circuit depth scale as the system size grows into regimes where classical computations become challenging. Since full classical simulation of the quantum algorithm is impractical for larger systems, we instead estimate the number of measurements from the analytical bounds as a function of the input parameters.

When employing the algorithm described in Sec.~\ref{sec:theory}, it is necessary to sample both the numerator and the denominator of Eq.~\eqref{eq:ITE} independently. Since these are independent sampling problems, we can distribute the total number of circuit samples between them to minimize the overall measurement budget.

Performing a Lagrangian optimization over this distribution (detailed in Appendix~\ref{app:optimization}), we obtain the following expression for the total optimal number of samples required to achieve an additive error $\varepsilon$ in the estimated expectation value:
\begin{align}
\label{eq:total_samples}
    M = \frac{\Gamma}{\varepsilon^2}\,,
\end{align}
where
\begin{align}
\label{eq:gamma}
    \Gamma < \frac{2\,R(\vec{r}\,)^4\,\ln(2/\delta)}{\gamma^4\,
    \mathrm{e}^{-2\lambda_0^2\tau^2}}\left(\|O\|+\left|\langle
    O\rangle_\mathrm{GS}\right|\right)^2\,.
\end{align}
We stress that Eq.~\eqref{eq:total_samples} gives the cost of estimating a single expectation value, i.e.\ one ground-state energy or one component of the force vector. Industrially relevant workflows require many such estimates: a single molecular dynamics step requires all $3N_a$ force components, and a converged trajectory typically requires $10^6$--$10^9$ steps, while geometry optimization and reaction-path following require $3N_a$ components at each of many configurations. The corresponding total cost is obtained by multiplying $M$ by the number of required expectation values, a factor we do not include in the estimates below.  It is further important to highlight wo features of Eq.~\eqref{eq:gamma} that dominate the cost . First, the total weight $R(\vec{r}\,)$ enters as $R^4$, so its minimization is essential. Second, the factor $e^{-2\lambda_0^2\tau^2}$ in the denominator means that the sample count grows as a Gaussian in the residual ground-state energy $E_0$ of the shifted Hamiltonian.

Here, $\delta$ is the failure probability (i.e., the algorithm returns a result within $\varepsilon$ of the true expectation value with probability at least $1-\delta$), $\gamma = |\langle \Phi_0  | \psi_0 \rangle|$ is the overlap between the initial state $|\psi_0\rangle$ and the ground state $|\Phi_0 \rangle$, $\|O\|$ denotes the operator norm of the observable $O$, and $\langle O \rangle_\mathrm{GS} = \langle \Phi_0  | O | \Phi_0  \rangle$ is the exact ground-state expectation value of the observable. In practice, an estimate of $\langle O \rangle_\mathrm{GS}$ may be obtained from classical methods such as DMRG or Hartree-Fock; if no estimate is available, one can use the bound $|\langle O \rangle_\mathrm{GS}| \leq \|O\|$. $\lambda_0 = E_0 - E_0^\mathrm{lb}$ is the residual ground-state energy of the shifted Hamiltonian $H' = H_0 + \Lambda\mathbbm{1}$ of Eq.~\eqref{eq:shift}. In the ideal case of an exact bound, $\lambda_0 = 0$ and the exponential factor equals unity; any $\lambda_0 > 0$ incurs a Gaussian penalty $e^{+2\lambda_0^2\tau^2}$ in the sample count. Note that $\lambda_0\tau$ is invariant under the normalization $\mathcal{N}$, so this penalty is unambiguous.  The quantity $R(\vec{r}\,)$ is the total weight of the combined Fourier-random-compiler decomposition, defined in Eq.~\ref{eq:random_unitaries}. Since $R(\vec{r}\,)$ enters the sample complexity as $R^4$, minimizing it is crucial for reducing the overall sampling complexity.

\subsection{Runtime parameters and evaluation of $R(r)$}
\label{subsec:runtime}

The runtime vector $\vec{r} = (r_1, \dots, r_{2N_z+1})$ controls the tradeoff between circuit depth and sample count: larger $r_k$ reduces $R(\vec{r}\,)$ but increases the number of non-Clifford gates per circuit. When using a single global $r$ for all Fourier terms (as in our numerical simulations of Sec.~\ref{sec:numerics}), $R$ can become very large for terms with large $|z_k|$.
A natural per-term choice is ~${\bf{r}}_\mathrm{analytical}$ with $r_k = \lambda^2 z_k^2 \tau^2$, where $\lambda$ is the Pauli weight of the Hamiltonian \cite{wang2024qubit}. With this choice, the random compiler weight for each Fourier term is bounded as $\sum_m |\beta_{km}^{(r_k)}| \leq e^{t_k^2 / r_k} = e$ (see Lemma~1 of Ref.~\cite{wang2024qubit}), giving $R(\vec{r}\,) \leq e \cdot \alpha$, where $\alpha = \sum_k |\alpha_k| \lesssim 1.4$ is the total weight of the Fourier coefficients~\cite{keen2021quantum}. This yields $R(\vec{r}\,) \lesssim 3.8$, which is essentially independent of system size and imaginary time.

Beyond this analytical choice, Ref.~\cite{wan2022randomized} introduces a numerical optimization procedure to find $\vec{r}_{\mathrm{opt}}$ minimizing the total cost $C_{\mathrm{tot}} = C_{\mathrm{sample}} \cdot C_{\mathrm{gate}}$, which balances sample count against circuit depth. In Appendix~\ref{appendix:r}, we describe how this optimization is adapted to RGPS.

For the per-term choice $r_k = \lambda^2 z_k^2 \tau^2$, the value of $R$ can be computed in closed form using the bound above. For arbitrary $r$ (including a single global value), we compute $R$ numerically by evaluating the sum
\begin{align}
    R = \sum_{k=-N_z}^{N_z} |\alpha_k| \left(\sum_{n\,\text{even}}
    q_n(t_k, r_k)\right)^{r_k},
\end{align}
where $q_n(t,r) = \frac{(t/r)^n}{n!}\sqrt{1 + (t/r/(n+1))^2}$ is the probability weight from the randomized compiler, $t_k = \lambda\, z_k \tau$ is the rescaled evolution time for the $k$-th Fourier term, and the sum is truncated at a sufficiently large even integer (we use $n_{\max} = 150$). For large systems where floating-point arithmetic overflows, we use arbitrary-precision decimal arithmetic.

\subsection{Input parameters for P450}
In the following subsections, we apply Eq.~\eqref{eq:gamma} to estimate the number of samples for an industrially relevant molecular system,  the active site of the cytochrome P450 bound to oxygen. This system has been studied previously  as a prototypical quantum computing use case in the pharmaceutical industry~\cite{goings2022reliably}. P450 plays an important role in the metabolism of drug molecules. In the catalytic cycle, one intermediate product is known as Compound~1, and consists of an Fe-porphyrin system bound to oxygen. To accurately predict reaction pathways, it is important to know the energetic ordering of the different spin states (doublet, quartet, or sextet). Determining this ordering is challenging for classical electronic structure methods, as demonstrated in Ref.~\cite{goings2022reliably}. 

We therefore estimate the number of samples required to determine the expectation value of the active-space Hamiltonian with respect to the approximate ground state, to a precision of $1$\,mHa. We note that estimating a different observable, such as the nuclear force, would be an equally relevant target; however, the resulting sample counts would be of a similar order of magnitude, since the dominant cost factors---overlap, gap, and imaginary time---are observable-independent. Moreover, the operator norms entering the bound are of comparable magnitude for the Hamiltonian and force operators, suggesting that the observable-dependent contribution does not qualitatively alter the scaling.
Finally, the force operators derived in Ref.~\cite{obrien2023purification} are defined in the full orbital space and do not directly apply to the active-space Hamiltonians considered here.

To evaluate $\Gamma$ in Eq.~\eqref{eq:gamma}, we need: (i)~a lower bound on the spectral gap $\Delta$, (ii)~the (lower bound on the) overlap $\gamma$ of the initial state with the ground state, (iii)~the Pauli weight $\lambda$ of the Hamiltonian, and (iv)~the residual ground-state energy $E_0$ after shifting. We estimate these parameters using DMRG calculations with Block2~\cite{zhai2023block2} on the active spaces from Ref.~\cite{goings2022data} for the spin-$5/2$ state of the P450-oxo Hamiltonian. The active spaces range from $11$ electrons in $8$ orbitals (active space~B) to $63$ electrons in $58$ orbitals (active space~X); see Table~\ref{tab:p450activespace} for details.

The spectral gaps are computed as the difference between the ground- and first-excited-state energies within the same spin sector, both obtained at bond dimension $4000$. The first excited state is found by searching for the lowest-energy state orthogonal to the ground state. To estimate the overlap $\gamma$, we compress the DMRG ground state to bond dimension~$1$ and compute the overlap of this compressed state with the original ground state. Unlike a Hartree-Fock overlap, which measures the quality of the specific Hartree-Fock determinant, this procedure identifies the single determinant with maximal overlap that can be obtained through the DMRG compression procedure. Consequently, the resulting overlap provides an upper bound on the overlap achievable by any single determinant initial state and therefore represents an optimistic estimate for $\gamma$ when using a single determinant on a quantum computer as initial state. The Pauli weight $\lambda$ is computed from the one- and two-electron integrals under the Jordan--Wigner transformation, excluding the identity contribution; since the observable is the shifted active-space Hamiltonian itself, we use $\lVert O\rVert \leq \lambda$ in Eq.~\eqref{eq:gamma}, so the reported sample counts are conservative. The imaginary time is reported in the dimensionless form $\lambda\tau$; divide by $\lambda$ from the same row to obtain $\tau$ in Ha$^{-1}$. The residual ground-state energy after shifting is taken as $E_0 = 10^{-3}$\,Ha.

From these parameters, we compute the imaginary time $\tau$ using Eq.~\eqref{eqn_appendix_tau} with a target infidelity of $\varepsilon_{\mathrm{state}} = 10^{-3}$. The resulting input parameters are shown in Tab.~\ref{tab:bound_parameters}.

For the C~active space ($19$ electrons, $15$ orbitals), it is feasible to verify the analytical $\tau$ upper bound by performing the imaginary time evolution $e^{-\tau^2 H'^2/2}|\psi_0\rangle$ directly using matrix product operator (MPO) algebra available in pyblock3~\cite{zhai2021pyblock3}. We proceed as follows.
First, we obtain the ground state of $H$ at bond dimension~$250$ using pyblock3 (which supports only $S_z$ symmetry, in contrast to Block2 which supports full SU(2) symmetry) and compress it to a single Slater determinant as the initial state. We then construct the MPO for $H$ at bond dimension~$200$ and for $H^2$ at bond dimension~$2000$; these values were chosen after a convergence study over different bond-dimension combinations for the MPO. To ensure a non-negative
spectrum, we shift the Hamiltonian by $\Lambda = 2756.512\,\mathrm{Ha}$, i.e.\ we take $E_0^\mathrm{lb} = -2756.512\,\mathrm{Ha}$, slightly below the DMRG ground-state energy ($E_0^{\mathrm{DMRG}} = -2756.511\,\mathrm{Ha}$ for this active space). The residual is then $\lambda_0 = E_0-E_0^\mathrm{lb} \approx 0.001\,\mathrm{Ha}$, matching the value assumed above, and the Gaussian filter $e^{-\tau^2 H'^2/2}$ preferentially preserves the ground-state component.

We evolve the compressed initial state under $H'^2$ in steps of $\Delta(\tau^2) = 0.05\,\mathrm{Ha}^{-2}$ up to $\tau^2 = 10\,\mathrm{Ha}^{-2}$, truncating the MPS to bond dimension~$250$ at each step. At every time step, we compute the overlap with the exact ground state. We fit an exponential decay $1 - |\langle \Phi_0 | \psi(\tau)\rangle| \sim a\, e^{-b\tau^2}$ to the deviation from unit overlap and extrapolate the imaginary time needed to achieve a fidelity of $0.999$. The resulting $\tau^2_{\mathrm{DMRG}} \approx 1090\,\mathrm{Ha}^{-2}$, i.e. $\tau_{\mathrm{DMRG}} \approx 33\,\mathrm{Ha}^{-1}$, which results in a dimensionless scaled value of $\lambda\tau_\mathrm{DMRG}=2155$; the analytical bound $\lambda\tau_{\mathrm{bound}} \approx 3104$, indicated by the star in Tab.~\ref{tab:bound_parameters}, therefore exceeds the DMRG estimate by roughly $44\%$. The analytical bound is thus reasonably tight for this system rather than grossly pessimistic.

\subsection{Resource estimates}

\begin{figure}[h]
\centering
\includegraphics[width=\columnwidth]{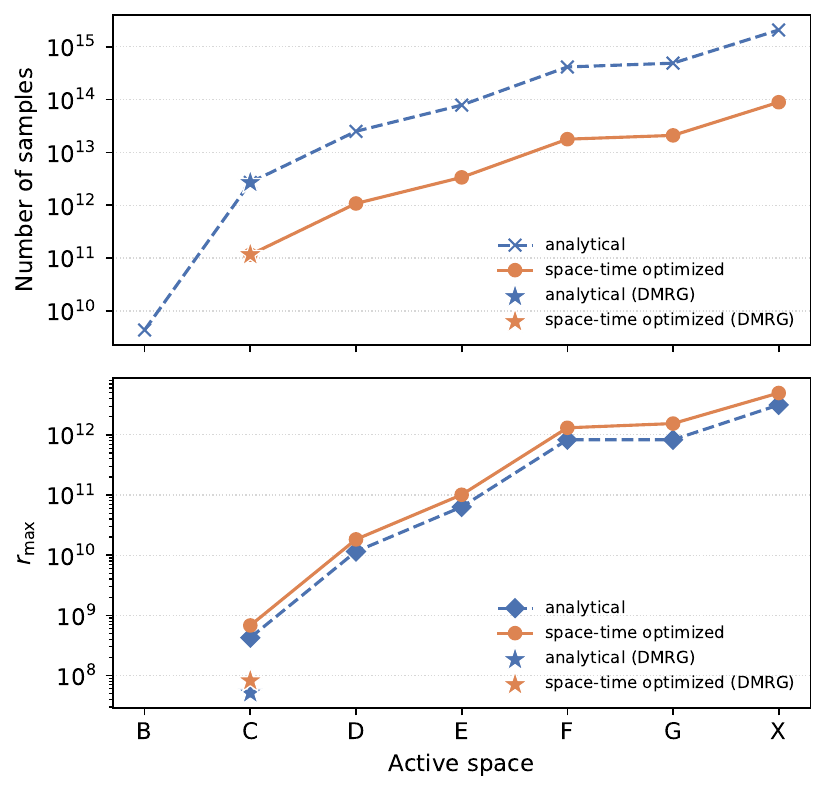}
\caption{\label{fig:P450_number_of_samples_analytical_r} Estimated number of samples required for the different P450 active spaces to estimate the Hamiltonian expectation value to an accuracy of $10^{-3}\,\mathrm{Ha}$. Results are shown for both the per-term runtime parameter $r_k = (\lambda\tau)^2 z_k^2$ and the space--time optimized runtime parameters described in Appendix~\ref{appendix:r}. Note that $\lambda\tau$ is the dimensionless imaginary time tabulated in Tab.~\ref{tab:bound_parameters}, so that $r_k$ follows directly from that column without further rescaling by $\lambda$. Compared to the analytical choice, the optimized parameters substantially reduce the maximum circuit depth while requiring only a modest increase in the number of samples. For active space~B, the overlap of the initial state with the target ground state already exceeds the target fidelity, resulting in $\tau=0$ and thus $r=0$. The stars correspond to calculations that use the imaginary-time evolution parameter obtained from the DMRG estimate for active space~C.}
\end{figure}

In Figure~\ref{fig:P450_number_of_samples_analytical_r}, we compare the estimated number of samples $M$ required to achieve chemical accuracy ($\varepsilon = 10^{-3}\,$Ha with failure probability $\delta = 0.01$) for the ground-state energy across all P450 active spaces. We consider two choices of the runtime vector as illustrated in Sec.~\ref{subsec:runtime} the analytical choice $\vec{r}_\mathrm{analytical}$, where $r_k = \lambda^2 z_k^2 \tau^2$ for each Fourier term~$k$ such that sampling weights are bounded to $\mathcal{O}(1)$ and the space-time optimized choice $\vec{r}_\mathrm{opt}$ that minimizes the total cost, circuit-depth and sampling-count of the entire RGPS procedure. Several trends are apparent. 
First, we note that the dominant cost for the smaller active spaces (e.g., B, where the overlap is nearly unity and no ground-state projection is required) is determined by the expectation-value estimation overhead, which scales as $\|O\|^2/\varepsilon^2$. For the B~active space with $\|O\| = \lambda_B \approx 24\,$Ha and $\varepsilon = 10^{-3}\,$Ha, this alone requires $\sim 10^{10}$ samples. Second, the sample count increases steeply with active-space size, driven primarily by the growth of the Pauli weight~$\lambda$ (which increases~$\tau$ through the rescaled gap~$\Delta/\lambda$) and the decrease in overlap~$\gamma$ (which enters as~$1/\gamma^4$). For the largest active space~X, the estimated sample count reaches $10^{15}$ with the analytical runtime vector, and $10^{14}$ with the space--time-optimized one.  In both cases, the maximal circuit depth across all Fourier points reaches~$\sim 10^{11}$ non-Clifford gates for active space~X. We note that these estimates are not optimized: the Hamiltonian is represented in the standard Pauli basis without applying factorization or compression techniques that could reduce the one norm $\lambda$. For comparison, standard QPE resources for this system are estimated to require $10^9$ Toffoli gates with $\mathcal{O}(1)$ repetitions.  Our estimates here could benefit also from low-rank factorizations such as tensor hypercontraction~\cite{lee2021even} or symmetry shifts~\cite{loaiza2023block, caesura2025faster}, but incorporating such compressions into the present randomized framework is not straightforward. 

\begin{table}[t]
    \centering
    \caption{
    Parameters entering the analytical sampling bound of Eq.~\eqref{eq:gamma} for the P450 active spaces. The ground-state energy $E_0$, spectral gap $\Delta$ and overlap $\gamma$ are obtained from DMRG calculations with a maximum bond dimension of 4000; the one-norm $\lambda$ is computed from the molecular integrals and the imaginary time $\tau$ from Eq.~\eqref{eqn_appendix_tau}. Here $E_0$ is the total ground-state energy of $H_0$; the residual energy after shifting is $\lambda_0=E_0-E_0^\mathrm{lb}=10^{-3}$\,Ha for all rows. Active space~B requires no imaginary-time evolution because the target fidelity is already achieved by the initial state. For the active space C, DMRG-based ITE was additionally used to estimate the required imaginary time, highlighted by the star.}
    \label{tab:bound_parameters}
    \vspace{1mm}
    \begin{tabular}{lccccr}
        \toprule
        Active Space &
        $E_0$ (Ha) &
        $\Delta$ (Ha) &
        $\gamma$ &
        $\lambda$ (Ha) &
        $\lambda\tau$  \\
        \midrule
        B        & -2756.493 & 0.0526 & 1.000 & 23.8   &      0      \\
        C        & -2756.511 & 0.0584 & 0.901 & 65.3   &   3104   \\
        C$^\ast$ & -2756.511 & 0.0584 & 0.901 & 65.3   &   2155   \\
        D        & -2756.539 & 0.0318 & 0.866 & 183.0  &  16115  \\
        E        & -2756.565 & 0.0226 & 0.842 & 303.1  &  37573  \\
        F        & -2756.720 & 0.0105 & 0.738 & 506.3  & 134441  \\
        G        & -2756.736 & 0.0104 & 0.733 & 541.7  & 145345  \\
        X        & -2756.945 & 0.0109 & 0.694 & 1006.3 & 261662 \\
        \bottomrule
    \end{tabular}
\end{table}

\begin{figure}[h]
\centering
\includegraphics[width=\columnwidth]{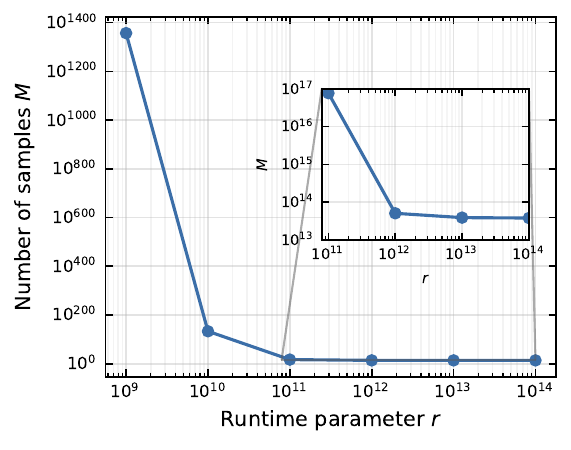}
\caption{\label{fig:P450_X_space} Estimated number of samples for estimating the expectation value of the Hamiltonian in the X~active space of P450, using a uniform $r$ parameter for all Fourier points. For small values of $r$, the required number of samples increases exponentially.}
\end{figure}

Specifically, an appealing feature of the randomized quantum algorithm is the ability to trade reduced circuit depth for an increased number of samples. In an early fault-tolerant setting with limited qubit counts and coherence times, only circuits up to some maximal depth~$r_\mathrm{max}$ may be executable. In this regime, the randomized algorithm remains runnable---at the cost of a larger even exponentially scaling  sample count---whereas the standard QPE algorithm may be infeasible entirely. To quantify this tradeoff, we consider a scenario in which all circuit samples use a uniform runtime parameter $r = r_\mathrm{max}$. Figure~\ref{fig:P450_X_space} shows the estimated number of samples~$M$ for a range of global~$r$ values from~$10^9$ to~$10^{14}$.

As expected, the choice of~$r$ significantly affects the estimate: smaller~$r$ leads to shallower circuits but exponentially more samples, while larger~$r$ brings the estimate closer to the theoretical minimum set by $R \approx e\,\alpha \lesssim 3.8$, as achieved with $\vec{r}_\mathrm{analytical}$. Overall, while the algorithm permits this tradeoff in principle, the number of samples quickly explodes into infeasible regimes for small~$r$, rendering the algorithm impractical in this setting.

\section{Conclusion}

In this work, we have benchmarked the randomized ground-state property sampling algorithm proposed by Wang et al.~\cite{wang2024qubit} for computing expectation values with respect to ground states of molecular systems. Through classical simulations on a hydrogen molecule and resource estimates for the industrially relevant cytochrome P450 system, we have quantified the measurement overhead and identified the key cost drivers for practical applications.

The randomized algorithm offers considerable flexibility and several appealing features: it enables the treatment of general matrix functions across a continuous range of circuit depths while requiring only a single auxiliary qubit. This depth-sample tradeoff is particularly well-suited to the early-fault-tolerant regime, where hardware resources remain constrained, and represents a qualitatively different approach compared to fully coherent algorithms.

Our analysis reveals that, for the molecular systems studied here, the measurement overhead remains a significant challenge. The algorithm permits arbitrarily shallow circuits in principle, but reducing the circuit depth below $r \sim (\lambda\tau)^2$ leads to a rapid increase in sampling complexity. For the systems considered, the required imaginary evolution time is of order$\lambda\tau \sim 10^3$--$10^5$, corresponding to maximal runtime parameters $r_{\max} \sim 10^8$--$10^{12}$. On top of this, the expectation-value estimation contributes an  overhead scaling as $\|O\|^2/\varepsilon^2$. For chemical accuracy ($\varepsilon \sim 10^{-3}$\,Ha) and operator norms $\|O\| \sim 10^2$--$10^3$\,Ha, this alone requires $10^{10}$--$10^{12}$ samples. This cost is clearly visible in the B~active space (Fig.~\ref{fig:P450_number_of_samples_analytical_r}), where the initial state already has near-unit overlap with the ground state, yet the sampling count remains $\sim 10^{10}$. This is in line with earlier resource estimates for linear system solvers based on randomized formulae~\cite{hariprakash2025randomized}. For the C~active space of P450, a direct DMRG simulation of the imaginary-time evolution allows the analytical bound on $\tau$ to be tested against the imaginary time actually required. The bound exceeds the DMRG estimate by roughly $44\%$, indicating that it is reasonably tight rather than grossly conservative. Since $\tau$ enters the circuit depth quadratically, sharpening this bound would reduce the runtime parameter by no more than a factor of about two, and we therefore do not expect improved bounds on $\tau$ to change the resource estimates qualitatively. We note that the DMRG value is itself obtained by extrapolation well beyond the directly simulated range, so the residual discrepancy is comparable to the uncertainty of that extrapolation.

Our results highlight several promising directions for future work. Most significantly, achieving Heisenberg-limited sample complexity---as opposed to the current shot-noise-limited scaling---would substantially reduce the measurement overhead and represents an important target for algorithmic development. Tighter analytical bounds on the required imaginary evolution time, improved Hamiltonian decomposition techniques that reduce the Pauli weight~$\lambda$, and the development of rigorous methods for bounding spectral gaps and initial-state overlaps would each contribute to closing the gap between theoretical promise and practical feasibility. Taken together, these advances could position randomized early-fault-tolerant algorithms as a viable route to industrially relevant quantum chemistry.

\section*{Acknowledgments}
We thank Mario Berta and Samson Wang for their valuable feedback on this manuscript. Additionally, we appreciate the members of the QUTAC Material Science working group for insightful discussions.

\bibliography{biblio}

\appendix
\section{Optimal distribution of shot count}
\label{app:optimization}

As detailed in Appendix~B of Ref.~\cite{wang2024qubit}, when sampling the
numerator, all possible outcomes will lie within the interval $[-\norm{O} R(\vec r)^2,\norm{O}
R(\vec r)^2]$. Following Eq.~(B16) in Ref.~\cite{wang2024qubit}, the number of
samples required to ensure the mean value of collected samples is within
$\varepsilon$ of the exact value, $|\bar{m}_{\mathrm{O}}-\mathrm{Tr}(s(H)\rho
s(H)O)|<\varepsilon_\mathrm{nom.}$, with failure probability $\delta$, is given
by
\begin{align}
    M_\mathrm{nom.}>\ln(2/\delta)\frac{2\norm{O}^2R(\vec r)^4}{\varepsilon_\mathrm{nom.}^2}\,.
    \label{eq:M_nom}
\end{align}

The same argument applies to the denominator, resulting in a sample count of
\begin{align}
    M_\mathrm{denom.}>\ln(2/\delta)\frac{2R(\vec r)^4}{\varepsilon_\mathrm{denom.}^2}\,.
    \label{eq:M_denom}
\end{align}
Since numerator and denominator are estimated independently, a union bound
implies that the combined estimator fails with probability at most $2\delta$; we
absorb this factor of two into $\delta$ and do not track it separately below.
Given errors of $\varepsilon_\mathrm{denom.}$ on the denominator and
$\varepsilon_\mathrm{nom.}$ on the numerator, Gaussian error propagation allows us
to compute the error on the estimate of the expectation value
$\bar{O}=\bar{m}_{\mathrm{nom.}}/\bar{m}_{\mathrm{denom.}}$ to
\begin{align}
    \varepsilon_\mathrm{O}^2&=\frac{\mathrm{nom.}^2}{\mathrm{denom.}^2}\left(\frac{\varepsilon_{\mathrm{nom.}}^2}{\mathrm{nom.}^2}+\frac{\varepsilon_{\mathrm{denom.}}^2}{\mathrm{denom.}^2} \right) 
    \label{eq:epsilon_O}
\end{align}

Substituting Eq.~\eqref{eq:M_nom} and Eq.~\eqref{eq:M_denom} into
Eq.~\eqref{eq:epsilon_O}, we obtain
\begin{align}
    \varepsilon_\mathrm{O}^2&=\frac{\mathrm{nom.}^2}{\mathrm{denom.}^2}\left(\frac{2\ln(2/\delta)\norm{O}^2R(\vec r)^4}{M_\mathrm{nom.}\,\mathrm{nom.}^2}+\frac{2\ln(2/\delta)R(\vec r)^4}{M_\mathrm{denom.}\,\mathrm{denom.}^2} \right)\nonumber\\
    &= 2R(\vec r)^4\ln(2/\delta)\frac{\mathrm{nom.}^2}{\mathrm{denom.}^2}\nonumber\\
    &\phantom{=}\times\left(\frac{\norm{O}^2}{M_\mathrm{nom.}\,\mathrm{nom.}^2}+\frac{1}{M_\mathrm{denom.}\,\mathrm{denom.}^2} \right)
\end{align}
To minimize the total shot count 
\begin{align}
    M=M_\mathrm{denom.}+M_\mathrm{nom.}\,,
\end{align}
we follow Eq.~(56) of~\cite{obrien2022efficient} and use Lagrangian optimization,
which for a constraint of the form $a/M_\mathrm{nom.}+b/M_\mathrm{denom.}=c$
yields $M=(\sqrt{a}+\sqrt{b})^2/c$ and hence an optimal shot count of
\begin{align}
    M=\frac{\Gamma}{\varepsilon^2}\,,
\end{align}
where $\Gamma$ defines the scaling factor 
\begin{align}
    \Gamma &= \frac{2R(\vec r)^4\ln(2/\delta)}{\mathrm{denom.}^2}\left(\norm{O}+\abs{\frac{\mathrm{nom.}}{\mathrm{denom.}}}\right)^2\,.
\end{align}
We further know that the denominator can be bound by
\begin{align}
    \mathrm{denom.}>\gamma^2 \mathrm{e}^{-\lambda_0^2\tau^2}\,,
\end{align}
with $\lambda_0=E_0-E_0^\mathrm{lb}$ the residual ground-state energy of the
shifted Hamiltonian defined in Eq.~\eqref{eq:residual}.

Assuming further that we have an estimate of the expectation value, so that
$\mathrm{nom.}/\mathrm{denom.}\approx\braket{O}_\mathrm{GS}$, we can
determine all parameters in Eq.~\eqref{eq:epsilon_O} and obtain
\begin{align}
    \Gamma <\frac{2R(\vec r)^4\ln(2/\delta)}{\gamma^4 \mathrm{e}^{-2\lambda_0^2\tau^2}}\left(\norm{O}+\abs{\braket{O}_\mathrm{GS}}\right)^2\,,
\end{align}
which is Eq.~\eqref{eq:gamma} of the main text.

\section{Cost reduction through optimized run time parameters}\label{appendix:r}

Wan et al.\ introduced in Ref.~\cite{wan2022randomized} a heuristic method to obtain adaptively chosen $\vec r=\{r_k\}_{k=1,\dots,N_k}$ parameters to obtain the global optimum regarding the total cost
\begin{equation}
    C_\mathrm{tot} = C_\mathrm{sample} \cdot C_\mathrm{gate}
\end{equation} where
\begin{align}
    C_\mathrm{sample} &\propto R(\vec r)^4 \\
    C_\mathrm{gate} &\propto \frac{1}{R(\vec r)} \sum_k |\alpha_k| u_k r_k.
\end{align}
$C_\mathrm{gate}$ is given as the average $r_k$, which is the count of non-Clifford operations per executed circuit, and $u_k=\exp(t_k^2/r_k)$ is the upper bound on the total randomized compilation weights. Ref.~\cite{wan2022randomized} applied the randomized compilation to a different algorithm, with a $C_\mathrm{sample}$ scaling only quadratically with the total weight $R(\vec r) = \sum_k |\alpha_k| u_k$, whereas RGPS has a quartic scaling, which necessitates a small adaptation. 
First, we define the approximated total cost 
\begin{equation}
    c(\vec r) = R(\vec r)^4 \cdot \frac{1}{R(\vec r)} \sum_k |\alpha_k| u_k r_k
\end{equation}
We want to find optimal choices for $\vec r$:
\begin{align}
    \frac{\partial c}{\partial r_l} &= -3 R(\vec r)^2 \frac{t_l^2}{r_l^2} u_l |\alpha_l| \sum_k |\alpha_k| u_k r_k \nonumber\\
    &\quad + R(\vec r)^3\left(1-\frac{t_l^2}{r_l}\right) u_l |\alpha_l| \\
    &= u_l|\alpha_l|R(\vec r)^3\left(1-\frac{t_l^2}{r_l}-3\frac{t_l^2}{r_l^2}S(\vec r)\right),
\end{align}
where
\begin{equation}
    S(\vec r) = \frac{\sum_k |\alpha_k|u_kr_k}{R(\vec r)}.
\end{equation}
Then, $\frac{\partial c}{\partial r_l} = 0$ if
\begin{equation}
    1-\frac{t_l^2}{r_l}-3\frac{t_l^2}{r_l^2}S(\vec r) = 0.
\end{equation}
Due to the additional factor $3$ compared to Ref.~\cite{wan2022randomized} (below Eq.~D7), we obtain, instead of the expression above Eq.~D8 of Ref.~\cite{wan2022randomized}, the fixed-point map
\begin{equation}
    \varrho_k(s) = \frac{t_k^2}{2}\left(1+\sqrt{1+12 s/t_k^2}\right),
\end{equation}
which we denote by $\varrho_k$ rather than $R_k$ to avoid a clash with the total weight $R(\vec r)$. Besides this modification, we can use the same numerical root-finding methods as proposed in Ref.~\cite{wan2022randomized}, Appendix D, to obtain optimal $r_k$'s.

\section{Parameter estimates for the ITE}\label{sec_param_estimate}
In this Appendix we present a short summary of parameters to set in numerical calculations (for the details see Appendix A of Ref.~\cite{keen2021quantum}).
\begin{itemize}
  \item Overlap of the ground state $\ket{\Phi_0}$ with the initial state $\ket{\psi_0}$: $\gamma = \abs{\braket{\Phi_0 | \psi_0}}$
  \item Imaginary time  $\tau_s$ such that  $\abs{\braket{\Phi_0|\psi}} \geq  1 - \varepsilon_\text{state}$ where $\ket{\psi} = f(H)\ket{\psi_0} / \sqrt{\braket{\psi_0 | f(H)^\dag
f(H) |\psi_0}} $: 
  \begin{equation}\label{eqn_appendix_tau}
  \tau_s \;\geq\; \frac{1}{\Delta_s}
  \sqrt{\frac{\log\!\left(\dfrac{1-\gamma^{2}}{\gamma^{2}\eta^{2}}\right)}
             {1 + 2\lambda_{0}^{(s)}/\Delta_{s}}}\,,
  \qquad \eta = \frac{\sqrt{2\,\varepsilon_\mathrm{state}}}{5}\,,
\end{equation}
where $\Delta_s=\Delta/\mathcal{N}$ and $\lambda_0^{(s)}=\lambda_0/\mathcal{N}$ are the spectral gap and the residual ground-state energy of the Hamiltonian rescaled by $\mathcal{N}=2\lambda$ (see Sec.~\ref{sec:theory}), and $\log$ denotes the natural logarithm. The physical imaginary time in Ha$^{-1}$ follows as $\tau=\tau_s/\mathcal{N}$, so that the dimensionless quantity tabulated in Tab.~\ref{tab:bound_parameters} is $\lambda\tau=\tau_s/2$. Note that $2\lambda_0^{(s)}/\Delta_s = 2\lambda_0/\Delta$, so the second factor is independent of $\mathcal{N}$; the same holds for the combination $\lambda_0^{(s)}\tau_s=\lambda_0\tau$.
\item In the limit $\lambda_0^{(s)}\to0$, and bounding the distance directly by
\begin{equation} 
    \norm{\ket{\psi} - \ket{\Phi_0}} \leq \tilde\eta = \sqrt{2\varepsilon_\text{state}}
\end{equation}
rather than by $\eta=\sqrt{2\varepsilon_\mathrm{state}}/5$, which additionally accounts for the truncation of the Fourier series, one recovers the simpler expression
\begin{equation}
    \tau_s \geq \frac{1}{\Delta_s}
     \left[2\log\left(\frac{1}{\gamma \tilde\eta }\right) + \log(1-\gamma^2)\right]^{1/2}\,.
\end{equation}
  \item Discretization step $\Delta_z$ and number of steps $N_z$ in the Fourier series 
  \begin{equation}
    \Delta_z = \frac{2\pi}{z'_c + \tau_s} \,,\qquad N_z = \left\lceil{z_c /  \Delta_z}\right\rceil\,,
  \end{equation}
  where 
  \begin{equation} 
    z_c =  \sqrt{2\log\dfrac{2}{\eta \gamma}} =  \sqrt{\log\frac{2}{\varepsilon_\text{state} \gamma^2}}\, ,
\end{equation}    
    and
\begin{equation}     
    z_c' =  [z_c^{2} + \arccosh (\pi z_c^2/4 + 1)]^{1/2}.
\end{equation}
\end{itemize}

\section{Specification of P450 active spaces}

\begin{table}[H]
    \centering
    \begin{tabular}{c|ccccccc}
        Active space  & B & C & D & E & F & G & X \\
        \hline
         elec.  & 11 & 19 & 25 & 33 & 45 & 47 & 63 \\
         orb.  & 8 & 15 & 23 & 31 & 41 & 43 & 58 \\
    \end{tabular}
    
    \caption{The active spaces of P450 utilized in this work are the same as the ``Cpd1'' system in Ref.~\cite{goings2022reliably}.}
    \label{tab:p450activespace}
\end{table}
\end{document}